\documentclass[11pt]{article}

\usepackage[a4paper,margin=2.5cm]{geometry}
\usepackage[T1]{fontenc}
\usepackage[utf8]{inputenc}

\usepackage{graphicx}
\usepackage{amsmath,amssymb,mathtools,bm}
\usepackage{booktabs,tabularx}
\usepackage{subcaption}
\usepackage[authoryear]{natbib}
\usepackage{authblk}

\newcommand{\Pran}{\mathrm{Pr}}
\newcommand{\Rey}{\mathrm{Re}}
\newcommand{\Ra}{\mathrm{Ra}}
\newcommand{\Rr}{\mathrm{Rr}}
\newcommand{\Ret}{\mathrm{Re}_{\tau}}
\newcommand{\Res}{\mathrm{Re}_{s}}
\newcommand{\Nu}{\mathrm{Nu}}

\newcommand{\ihc}{\mathrm{IHC}}
\newcommand{\Dt}{\widetilde{\Delta}}
\newcommand{\Qz}{\widetilde{Q}_0}
\newcommand{\epsu}{\epsilon_u}
\newcommand{\epst}{\epsilon_{\theta}}
\newcommand{\order}[1]{\mathcal{O}(#1)}

\title{Ultimate regimes in horizontal and internally heated convection}

\author[1]{Olga Shishkina}
\author[1,2]{Detlef Lohse}

\affil[1]{Max Planck Institute for Dynamics and Self-Organization, G\"{o}ttingen, 37077, Germany}
\affil[2]{Physics of Fluids Department, Faculty of Science and Technology, University of Twente,
7500, AE Enschede, The Netherlands}

\date{}

\begin{document}

\maketitle

\begin{center}
\small
\texttt{\texttt{Olga.Shishkina@ds.mpg.de}}\\
\texttt{\texttt{d.lohse@utwente.nl}}
\end{center}

\begin{abstract}
We derive asymptotic models for the ultimate regimes in horizontal convection (HC) and pure internally heated convection (IHC), in analogy with our recent (2024) extension of the ultimate-regime model for Rayleigh--B\'enard convection (RBC).
To derive the corresponding models for HC and IHC, we combine turbulent boundary-layer relations with the exact dissipation balances for these two systems.
For HC, the resulting scaling relations are consistent with the rigorous transport bound of \citet{Siggers2004}.
For pure IHC, they are consistent with the exact HC--IHC balance analogy of \citet{Wang2021} and with the rigorous bounds on the convective-flux asymmetry in the equal-temperature-plates configuration \citep{Arslan2021}.
The main difference between RBC and HC/IHC is that, in the latter two cases, the global kinetic-energy balance does not contain the additional response factor
(dimensionless convective heat flux in HC or inversed bulk temperature in IHC), whereas it does in RBC.
As a consequence, for fixed $\Pran$, the ultimate-regime scaling exponent is $1/3$ for both HC and IHC, rather than $1/2$ as in RBC.
\end{abstract}

\section{Introduction}
The ultimate regime of thermal convection corresponds to very strong thermal driving, in which the boundary layers undergo a transition from laminar to turbulent, and the global heat-transport scaling laws change.
For Rayleigh--B\'enard convection (RBC), this problem has been central for decades, because RBC is the canonical model system for turbulent heat transport and because extrapolations to geo- and astrophysical flows depend on these asymptotic scaling laws \citep{Ahlers2009, Lohse2024}.

The classical regime of RBC is quantitatively described by the Grossmann--Lohse (GL) theory \citep{Grossmann2000,Grossmann2001}. The ultimate regime, however, remained controversial for a long time, both experimentally and theoretically \citep{Ahlers2009,Roche2010,Lohse2023,Lohse2024}.
The model of \citet{Shishkina2024}, which may be viewed as an extension of the \citet{Grossmann2011} model of the ultimate regime to the high-Prandtl-number regime ($\Pran \gg 1$) in RBC, provides asymptotic scaling predictions that are consistent with the mathematically rigorous upper bounds on heat transport. This was not the case for earlier ultimate-regime models \citep{Kraichnan1962,Spiegel1971,Chavanne1997,Grossmann2011}.

These bounds on the relation between the dimensionless heat transport (Nusselt number $\Nu$) and the thermal driving (Rayleigh number $\Ra$) are essential.
For no-slip RBC they imply $\Nu \lesssim \Ra^{1/2}$ \citep{Howard1963,Doering1996,Seis2015}, which already excludes asymptotic laws such as $\Nu \sim \Pran^{1/2}\Ra^{1/2}$ when $\Pran$ increases with $\Ra$.
In the very large-$\Pran$ regime, a sharper bound further implies
$\Nu \lesssim \Ra^{1/3}$,
up to logarithmic corrections \citep{Constantin1999,Choffrut2016}; see also the further discussion in \citet{Lohse2024}.

Horizontal convection (HC), a configuration relevant to large-scale geophysical flows \citep{Rossby1965,Hughes2008}, is a natural next system in which to investigate the ultimate regime.
In HC, heating and cooling are imposed on different parts of the same horizontal boundary.
Previous HC studies established Rossby's laminar scaling and its later refinements, while also showing that the flow becomes increasingly complex as $\Ra$ increases \citep{Paparella2002, Mullarney2004, Scotti2011}.
For large-aspect-ratio containers, \citet{Shishkina2016} derived the laminar low-$\Pran$ and high-$\Pran$ branches and confirmed them numerically.
Based on analytical relations for the HC dissipation balances and laminar boundary-layer estimates, \citet{Shishkina2016a} proposed a GL-type scaling model for the classical regime in HC.
Subsequent studies analysed the mean-flow structure in large-aspect-ratio HC and discussed a possible route towards turbulent HC \citep{Shishkina2017b, Tsai2016, Tsai2020, Reiter2020, Passaggia2024, Passaggia2024a}.
Not less important is the rigorous work by \citet{Siggers2004}, who derived an upper bound with the scaling exponent $1/3$ in the Nusselt-vs.-Rayleigh number dependence.
This exponent (and not the RBC exponent $1/2$) is the mathematically exact asymptotic upper limit to be respected in HC.

Pure internally heated convection (IHC) provides a second fundamental extension of RBC.
In the canonical wall-bounded IHC models, a fluid layer is heated volumetrically while the plates are either both isothermal and kept at the same temperature, or the top plate is isothermal and the bottom one insulating \citep{Goluskin2016b}.
Our work addresses the first of these setups, namely pure IHC with equal-temperature top and bottom plates.
In this configuration the buoyancy forcing is generated in the bulk rather than by an imposed wall-to-wall temperature difference, and the relevant global responses are the dimensionless mean bulk temperature $\Dt$ and the Reynolds number $\Rey$ \citep{Wang2021}.
A GL-type scaling theory for this pure-IHC setup was derived in \citet{Wang2021}, where the classical-regime branches for $\Dt$ and $\Rey$ were obtained and the formal HC--IHC analogy at the level of exact balances and regime structure was emphasized.

A different line of IHC research concerns internally heated and internally cooled source-sink systems, in which part or all of the generated heat is absorbed again in the bulk before reaching a wall \citep{Lepot2018, Bouillaut2019, Miquel2020, Kazemi2022}.
In such systems, mixing-length-type transport with a $1/2$ exponent can occur because transport between source and sink regions can bypass the wall boundary layers.
In particular, \citet{Kazemi2022} studied a top-isothermal/bottom-insulating layer with an imposed heating-cooling profile and found a continuous change of the effective heat-transport exponent from about $1/3$ to about $1/2$ as heating and cooling were made more balanced.
These results, however, do not describe the pure-IHC setup considered here, in which the net internally generated heat must still leave through the boundaries.

In the present study, we derive asymptotic models for the ultimate regimes of HC and pure IHC, in the spirit of the analogous RBC derivation of \citet{Shishkina2024}.
The turbulent boundary-layer relations are retained, and only the exact global kinetic-energy balances are replaced by the forms relevant to HC and IHC.
This single modification has a major consequence: for fixed $\Pran$, the ultimate-regime scaling exponent becomes $1/3$ instead of $1/2$.
We also derive all $\Pran$-dependent  ultimate subregimes and the slopes of the transition ranges.

\section{Wall-bounded turbulent flows}

We now recall only those elements of the derivation by \citet{Shishkina2024} that are needed below.
Close to a heated or cooled no-slip wall, the time- and area-averaged ($\langle \cdot \rangle$) momentum and temperature equations reduce to
\begin{equation}
\partial_z\langle u_z' u_x'\rangle = \nu\,\partial_z^2\langle u_x\rangle,
\qquad
\partial_z\langle u_z' \theta'\rangle = \kappa\,\partial_z^2\langle \theta\rangle.
\label{eq:reducedBL}
\end{equation}
Here $u_x$ and $u_z$ denote the horizontal and vertical velocity components, respectively, while $u_x'$ and $u_z'$ are their fluctuations; $\theta$ is the temperature;
$\nu$ is the kinematic viscosity and $\kappa$ the thermal diffusivity.
With the eddy viscosity $\nu_\tau$ and the eddy thermal diffusivity $\kappa_\tau$ defined by
$\langle u_z' u_x'\rangle \equiv -\nu_\tau\,\partial_z\langle u_x\rangle$,
and
$\langle u_z' \theta'\rangle \equiv -\kappa_\tau\,\partial_z\langle \theta\rangle$,
for the friction velocity $u_\tau\equiv\sqrt{\nu\partial_z\left.\langle u_x\rangle\right|_{z=0}}$
and the Nusselt number
$\Nu\equiv-(L/\Delta)\partial_z\left.\langle\theta\rangle\right|_{z=0}$, $\Delta\equiv\theta_+-\theta_-$,
where $\theta_+$ ($\theta_-$) is the temperature of the heated (cooled) plate,
we obtain
\begin{equation}
 u_\tau^2 = (\nu+\nu_\tau)\,\partial_z\langle u_x\rangle,
 \qquad
 ({\kappa\Delta}/{L})\Nu = -(\kappa+\kappa_\tau)\,\partial_z\langle\theta\rangle.
\label{eq:wallflux}
\end{equation}
Here $L$ is the reference length scale: the height of the domain in RBC and IHC, and the horizontal length of the domain in HC.

Outside the viscous sublayer, we assume a Landau-type closure
\begin{equation}
\nu_\tau \sim u_\tau z\Pran^{\zeta},
\qquad
\kappa_\tau \sim u_\tau z\Pran^{\zeta},
\qquad
\epsu(z)\sim \frac{u_\tau^3}{z\Pran^{\zeta}},
\label{eq:landau}
\end{equation}
with $z$ being the distance to the wall and
\begin{equation}
\zeta = 0 \quad (\Pran\lesssim 1),
\qquad
\zeta = -1/2 \quad (\Pran\gtrsim 1).
\label{eq:zeta}
\end{equation}
Integrating across the turbulent boundary region yields the generic relations
\begin{equation}
\Nu \sim \frac{\Pran^{\zeta+1}\Ret}{\log \Ret},
\qquad
\Rey \sim \Pran^{-\zeta}\Ret\log \Ret,
\label{eq:ReNuRet}
\end{equation}
and the corresponding bulk kinetic dissipation estimate
\begin{equation}
\epsu \sim ({\nu^3}/{L^4})\Pran^{-\zeta}\Ret^3\log \Ret,
\label{eq:epsRet}
\end{equation}
where $\Ret \equiv u_\tau L/\nu$.

In RBC, these expressions are combined with the exact balance
\begin{equation}
\epsu = ({\nu^3}/{L^4})\Pran^{-2}(\Nu-1)\Ra,
\label{eq:RBCexact}
\end{equation}
which involves the Rayleigh number
$\Ra \equiv {\alpha g \Delta L^3}/({\nu\kappa})$
and the Prandtl number
$\Pran \equiv {\nu}/{\kappa}$,
where $g$ is the gravitational acceleration and $\alpha$ is the thermal expansion coefficient.
This leads to the ultimate-regime scaling exponent $1/2$.
In HC and IHC, we retain \eqref{eq:ReNuRet} and \eqref{eq:epsRet}, but replace \eqref{eq:RBCexact} by the corresponding exact balances for HC and pure IHC.

\section{Horizontal convection}

\subsection{Classical regime and rigorous constraints}
We consider large-aspect-ratio HC, in which heating and cooling are applied to different parts of the same horizontal boundary, while all other walls are no-slip and adiabatic.
The control parameters are the Rayleigh ($\Ra$) and Prandtl ($\Pran$) numbers,
and the responses are the Nusselt ($\Nu$) and Reynolds ($\Rey$) numbers.
The classical GL-type HC model is built on the dissipation balances \citep{Shishkina2016a}
\begin{equation}
\epst \sim ({\kappa\Delta^2}/{L^2})\Nu,
\qquad
\epsu \sim ({\nu^3}/{L^4})\Ra\Pran^{-2}.
\label{eq:HCineq}
\end{equation}
The crucial point is the absence of the extra factor $\Nu$ in the kinetic balance.
This difference from RBC explains why the HC exponents are smaller than in RBC for the same assumed dissipation mechanism.

The GL-model \citep{Shishkina2016a}, together with the DNS studies \citep{Shishkina2016, Reiter2020}, yielded, among other branches, the well-established laminar low-$\Pran$ branch,
$\Nu \sim \Pran^{1/10}\Ra^{1/5}$ and $\Rey \sim \Pran^{-4/5}\Ra^{2/5}$,
as well as the large-$\Pran$ branch,
$\Nu \sim \Ra^{1/4}$ and $\Rey \sim \Pran^{-1}\Ra^{1/2}$.
The mean-flow study of \citet{Shishkina2017b} further showed, for large aspect ratio, how the mean temperature, velocity, and dissipation fields reorganize with increasing $\Ra$, and discussed a possible route towards a cell-wide turbulent HC state.

Concerning rigorous upper bounds for HC, \citet{Siggers2004} derived an upper bound on the horizontal heat transport using entropy production and a pseudo-flux formulation. For a general surface temperature distribution and lower-boundary conditions involving the temperature gradient, including insulating and constant-flux cases, they obtained
\begin{equation}
\Nu \lesssim \Ra^{1/3}.
\label{eq:Siggers13}
\end{equation}

\subsection{Ultimate branches}

In our model, we assume that the inequality in \eqref{eq:HCineq} is saturated at scaling level within the asymptotic regime,
\begin{equation}
\epsu \sim ({\nu^3}/{L^4})\Ra\Pran^{-2}.
\label{eq:HCexact}
\end{equation}
Substituting \eqref{eq:HCexact} into the generic turbulent boundary flow relation \eqref{eq:epsRet} gives
\begin{equation}
\Pran^{-2}\Ra \sim \Pran^{-\zeta}\Ret^3\log \Ret.
\label{eq:HCRet}
\end{equation}
Together with \eqref{eq:ReNuRet}, this leads to the generic ultimate HC scaling laws
\begin{equation}
\Rey \sim \Pran^{-(2+2\zeta)/3}\Ra^{1/3}[\log \Ra]^{2/3},
\qquad
\Nu \sim \Pran^{(1+4\zeta)/3}\Ra^{1/3}[\log \Ra]^{-4/3},
\label{eq:HCgeneric}
\end{equation}
where, exactly as in \citet{Shishkina2024} for the RBC case, we have replaced $\log \Ret$ by $\log \Ra$ at leading order and suppressed the $\Pran$ dependence inside the logarithms. Using \eqref{eq:zeta}, we obtain the two ultimate branches
\begin{align}
\mathrm{IV}'_{\ell}:\quad
&\Nu \sim \Pran^{1/3}\Ra^{1/3}[\log \Ra]^{-4/3},
&\Rey \sim \Pran^{-2/3}\Ra^{1/3}[\log \Ra]^{2/3},
&& \Pran\lesssim 1,
\label{eq:HCIVl}\\[1mm]
\mathrm{IV}'_{u}:\quad
&\Nu \sim \Pran^{-1/3}\Ra^{1/3}[\log \Ra]^{-4/3},
&\Rey \sim \Pran^{-1/3}\Ra^{1/3}[\log \Ra]^{2/3},
&& \Pran\gtrsim 1.
\label{eq:HCIVu}
\end{align}
These branches are the HC counterparts of the two ultimate RBC branches $\mathrm{IV}'_{\ell}$ and $\mathrm{IV}'_{u}$ from \cite{Shishkina2024}.
The main difference is the $\Ra$-exponent $1/3$, which is exactly the exponent imposed by the rigorous HC bound \eqref{eq:Siggers13}.

The adjoining ultimate subregimes follow from the same matching logic as in \citet{Shishkina2024}.
The two ultimate branches \eqref{eq:HCIVl} and \eqref{eq:HCIVu} meet at
$\Pran \sim \Ra^{0}$.

For small $\Pran$, let $\Pran\sim \Ra^{\eta}$ along a transition line. Then \eqref{eq:HCIVl} gives $\Nu\sim \Ra^{(1+\eta)/3}$ up to logarithmic corrections.
The limiting admissible slope is reached when the Nusselt number stops to grow, namely at $\eta=-1$, i.e.,
$\Pran \sim \Ra^{-1}$.
Along this line the ultimate branch matches the ultimate branch
\begin{equation}
\mathrm{II}'_{\ell}:\qquad
\Nu \sim \Pran^{1/6}\Ra^{1/6},
\qquad
\Rey \sim \Pran^{-2/3}\Ra^{1/3}.
\label{eq:HCIIl}
\end{equation}

For large $\Pran$, the relevant restriction is provided by the Friedrichs inequality \cite[see details in ][]{Shishkina2021}, which for no-slip walls gives
\begin{equation}
\Rey^2 \lesssim ({L^4}/{\nu^3})\epsu \sim \Pran^{-2}\Ra.
\label{eq:HCFriedrichs}
\end{equation}
Substituting \eqref{eq:HCIVu} into \eqref{eq:HCFriedrichs} yields, up to logarithmic corrections,
$\Pran \lesssim \Ra^{1/4}$.
Along the bounding slope $\Pran \sim \Ra^{1/4}$, \eqref{eq:HCIVu} gives $\Nu\sim \Ra^{1/4}$ and $\Rey\sim \Ra^{1/4}$, which matches the upper ultimate branch
\begin{equation}
\mathrm{III}'_{\infty}:\qquad
\Nu \sim \Ra^{1/4},
\qquad
\Rey \sim \Pran^{-1}\Ra^{1/2}.
\label{eq:HCIIIinf}
\end{equation}

The onset of the HC ultimate regime is estimated in the same way as in the classical GL picture, namely when the shear Reynolds number $\Res \sim \Rey^{1/2}$ reaches a sufficiently large value.
This implies an onset transition region of the form $\Pran \sim \Ra^{1/2}$.
The proposed HC ultimate regime therefore consists of the four subregimes $\mathrm{II}'_{\ell}$, $\mathrm{IV}'_{\ell}$, $\mathrm{IV}'_{u}$, and $\mathrm{III}'_{\infty}$, of which $\mathrm{IV}'_{\ell}$ and $\mathrm{IV}'_{u}$ are the two ultimate branches.

\section{Internally heated convection}

\subsection{Classical regime and rigorous constraints}

We consider pure internally heated convection between two no-slip plates kept at the same constant temperature.
This is the same setup as in \citet{Wang2021} and \citet{Arslan2021}.
In \citet{Wang2021} the horizontal directions are periodic; for the present scaling argument this lateral detail is secondary, because the exact balances and the asymptotic wall-layer relations are controlled by the two isothermal plates.
The control parameters are the Prandtl number $\Pran$ and the Rayleigh--Roberts number
\begin{equation}
\Rr = {\alpha g\Omega L^5}/({\kappa^2\nu}),
\end{equation}
where $\Omega$ is the volumetric heating rate.
The main response quantities are the Reynolds number $\Rey$ and the dimensionless mean temperature
\begin{equation}
\Dt = {\kappa\Theta}/({\Omega L^2}),
\end{equation}
with $\Theta$ the mean bulk temperature.

The exact dissipation balances from \citet{Wang2021} read
\begin{equation}
\epst = ({\kappa\Theta^2}/{L^2})\Dt^{-1},
\qquad
\epsu = ({\nu^3}/{L^4})\Rr\Pran^{-2}\Phi,
\label{eq:IHCexact}
\end{equation}
where $\Phi\equiv\langle u_z\theta\rangle_V$, $0\leq \Phi \leq 1/2$,  is the time- and volume-averaged vertical convective heat flux,
so that the outward heat fluxes through the top and bottom plates are, respectively, $1/2+\Phi$ and $1/2-\Phi$.
(In the notation of \citet{Wang2021}, $\Phi \equiv 1/2-\Qz$, where $\Qz$ is the dimensionless outward heat flux through the bottom plate.)

Treating $\Phi$ as order unity at scaling level, \citet{Wang2021} showed that the exact-balance structure of pure IHC is formally identical to that of HC with the mapping
\begin{equation}
\Ra \longleftrightarrow \Rr,
\qquad
\Nu \longleftrightarrow \Dt^{-1}.
\label{eq:HC_IHC_map}
\end{equation}
This transfers the classical HC phase diagram to pure IHC.
In particular, \citet{Wang2021} obtained the low-$\Pran$ classical branch
$\Dt \sim \Pran^{-1/10}\Rr^{-1/5}$,
$\Rey \sim \Pran^{-4/5}\Rr^{2/5}$,
the adjoining lower-$\Pran$ branch
$\Dt \sim \Pran^{-1/6}\Rr^{-1/6}$,
$\Rey \sim \Pran^{-2/3}\Rr^{1/3}$,
and the very-large-$\Pran$ branch
$\Dt \sim \Rr^{-1/4}$,
$\Rey \sim \Pran^{-1}\Rr^{1/2}$.

For the equal-temperature-plates IHC setup, the rigorous quantity studied so far is not $\Dt^{-1}$ but the mean vertical convective heat flux $\Phi$.
\citet{Arslan2021} proved $\langle \Phi\rangle\leq 1/2$ and obtained an explicit $\Rr$-dependent improvement $\langle wT\rangle\leq 2^{-21/5}\Rr^{1/5}$ over a finite range.
This bound is better than $1/2$ only up to $\Rr<2^{16}=65536$, but it is rigorous and demonstrates that the trivial upper bound can be sharpened over a finite parameter range \citep{Arslan2021}.
With the temperature minimum principle enforced, their numerically optimized bounds remain strictly below $1/2$ up to the largest $\Rr$ they considered and appear to approach $1/2$ from below.
In the present context, these results say only that the prefactor $\Phi$ in the exact kinetic balance \eqref{eq:IHCexact} is bounded and at most order unity.

\subsection{Utimate branches}
Under the scaling assumption \eqref{eq:IHCexact} with $\Phi=\order{1}$, the kinetic balance has the same structure as \eqref{eq:HCexact}, and the derivation of the ultimate IHC branches is algebraically identical to the HC one. Substituting \eqref{eq:IHCexact} into \eqref{eq:epsRet} and using \eqref{eq:ReNuRet} together with the map \eqref{eq:HC_IHC_map}, we obtain
\begin{equation}
\Rey \sim \Pran^{-(2+2\zeta)/3}\Rr^{1/3}[\log \Rr]^{2/3},
\qquad
\Dt^{-1} \sim \Pran^{(1+4\zeta)/3}\Rr^{1/3}[\log \Rr]^{-4/3}.
\label{eq:IHCgeneric}
\end{equation}
Hence the two ultimate IHC branches are
\begin{align}
\mathrm{IV}'_{\ell}:\qquad
&\Dt \sim \Pran^{-1/3}\Rr^{-1/3}[\log \Rr]^{4/3},
&\Rey \sim \Pran^{-2/3}\Rr^{1/3}[\log \Rr]^{2/3},
&& \Pran\lesssim 1,
\label{eq:IHCIVl}\\[1mm]
\mathrm{IV}'_{u}:\qquad
&\Dt \sim \Pran^{1/3}\Rr^{-1/3}[\log \Rr]^{4/3},
&\Rey \sim \Pran^{-1/3}\Rr^{1/3}[\log \Rr]^{2/3},
&& \Pran\gtrsim 1.
\label{eq:IHCIVu}
\end{align}
The inverse mean temperature therefore grows as $\Rr^{1/3}$ at fixed $\Pran$, exactly as suggested by the HC analogy and by the boundary-limited nature of purely IHC.

The adjoining subregimes again follow by matching.
The two ultimate branches meet at $\Pran\sim \Rr^{0}$.
For small $\Pran$, the limiting admissible slope is
$\Pran \sim \Rr^{-1}$,
which yields the lower-$\Pran$ branch
\begin{equation}
\mathrm{II}'_{\ell}:\qquad
\Dt \sim \Pran^{-1/6}\Rr^{-1/6},
\qquad
\Rey \sim \Pran^{-2/3}\Rr^{1/3}.
\label{eq:IHCIIl}
\end{equation}
For large $\Pran$, the Friedrichs inequality implies
$\Rey^2 \lesssim ({L^4}/{\nu^3})\epsu \sim \Rr\Pran^{-2}$,
so that, with \eqref{eq:IHCIVu},
$\Pran \lesssim \Rr^{1/4}$
up to logarithmic corrections.
This gives the upper-$\Pran$ branch
\begin{equation}
\mathrm{III}'_{\infty}:\qquad
\Dt \sim \Rr^{-1/4},
\qquad
\Rey \sim \Pran^{-1}\Rr^{1/2}.
\label{eq:IHCIIIinf}
\end{equation}
Finally, with $\Res\sim \Rey^{1/2}$ and $\Rr\Pran^{-2}\sim \Rey^3$, the onset of the IHC ultimate sector is
\begin{equation}
\Rr_{cr}^{(\ihc)} \sim \Pran^2\mathrm{Re}_{s,cr}^{6},
\qquad \text{equivalently} \qquad
\Pran \sim \Rr^{1/2}.
\label{eq:IHConset}
\end{equation}
Thus the proposed pure-IHC ultimate subregimes include $\mathrm{II}'_{\ell}$, $\mathrm{IV}'_{\ell}$, $\mathrm{IV}'_{u}$, and $\mathrm{III}'_{\infty}$, with $\mathrm{IV}'_{\ell}$ and $\mathrm{IV}'_{u}$ being the two ultimate branches.

\begin{figure}
\centering
\begin{subfigure}[t]{0.6\textwidth}
\centering
\includegraphics[width=\linewidth]{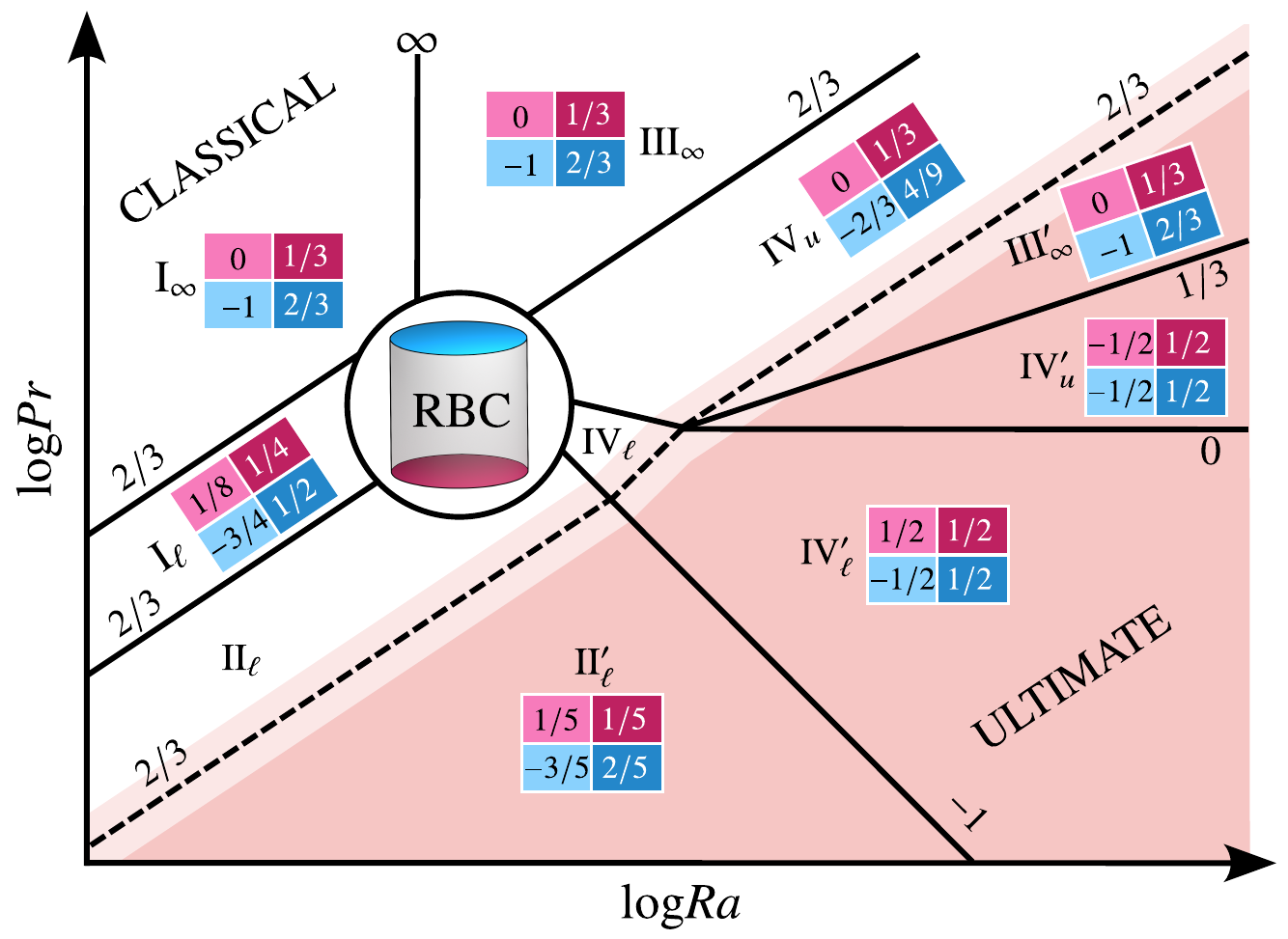}
\put(-240,160){{\fontsize{10pt}{12pt}\selectfont(\textit{a})}}
\end{subfigure}
\hfill
\begin{subfigure}[t]{0.6\textwidth}
\centering
\includegraphics[width=\linewidth]{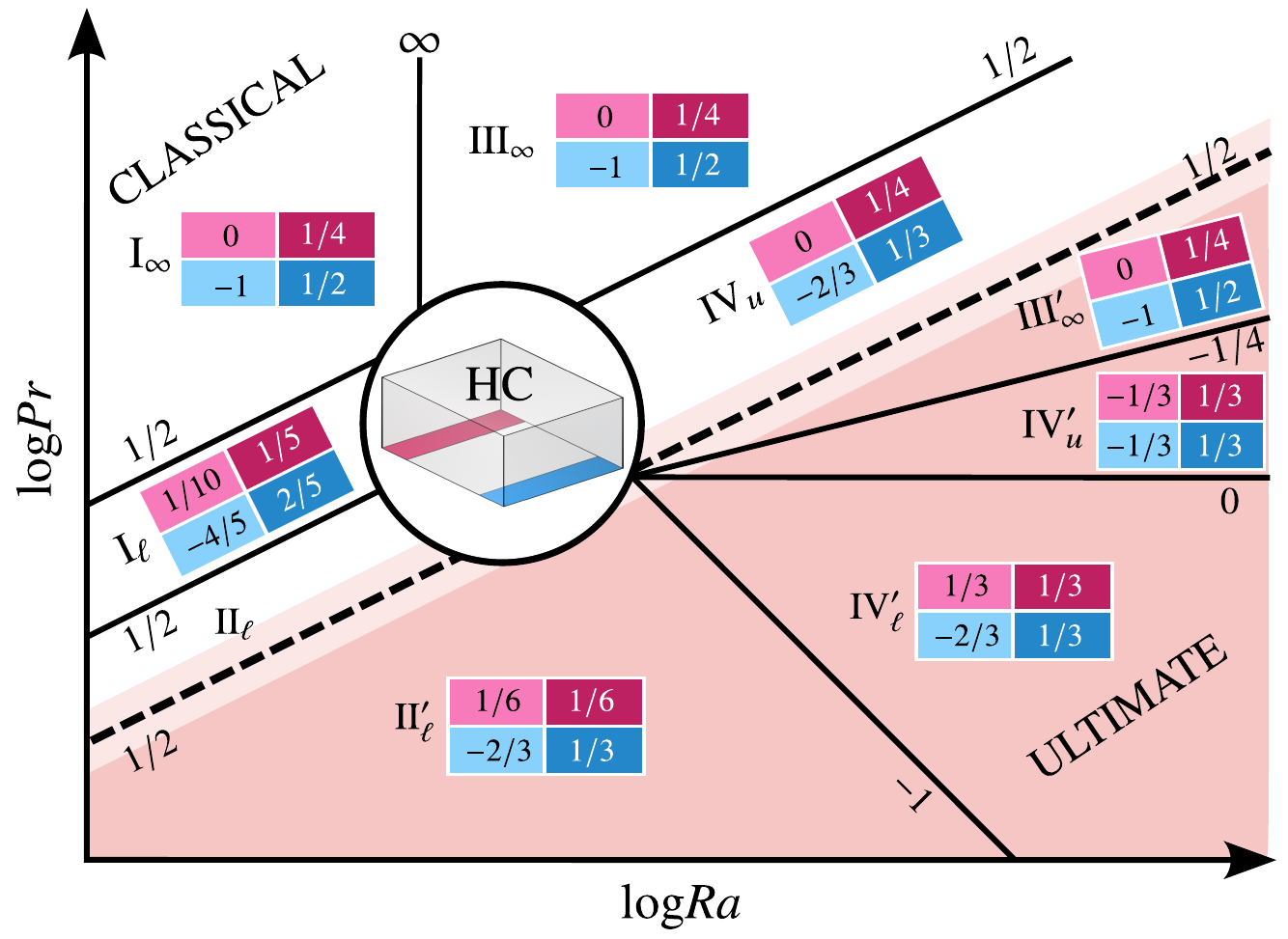}
\put(-240,160){{\fontsize{10pt}{12pt}\selectfont(\textit{b})}}
\end{subfigure}
\hfill
\begin{subfigure}[t]{0.6\textwidth}
\centering
\includegraphics[width=\linewidth]{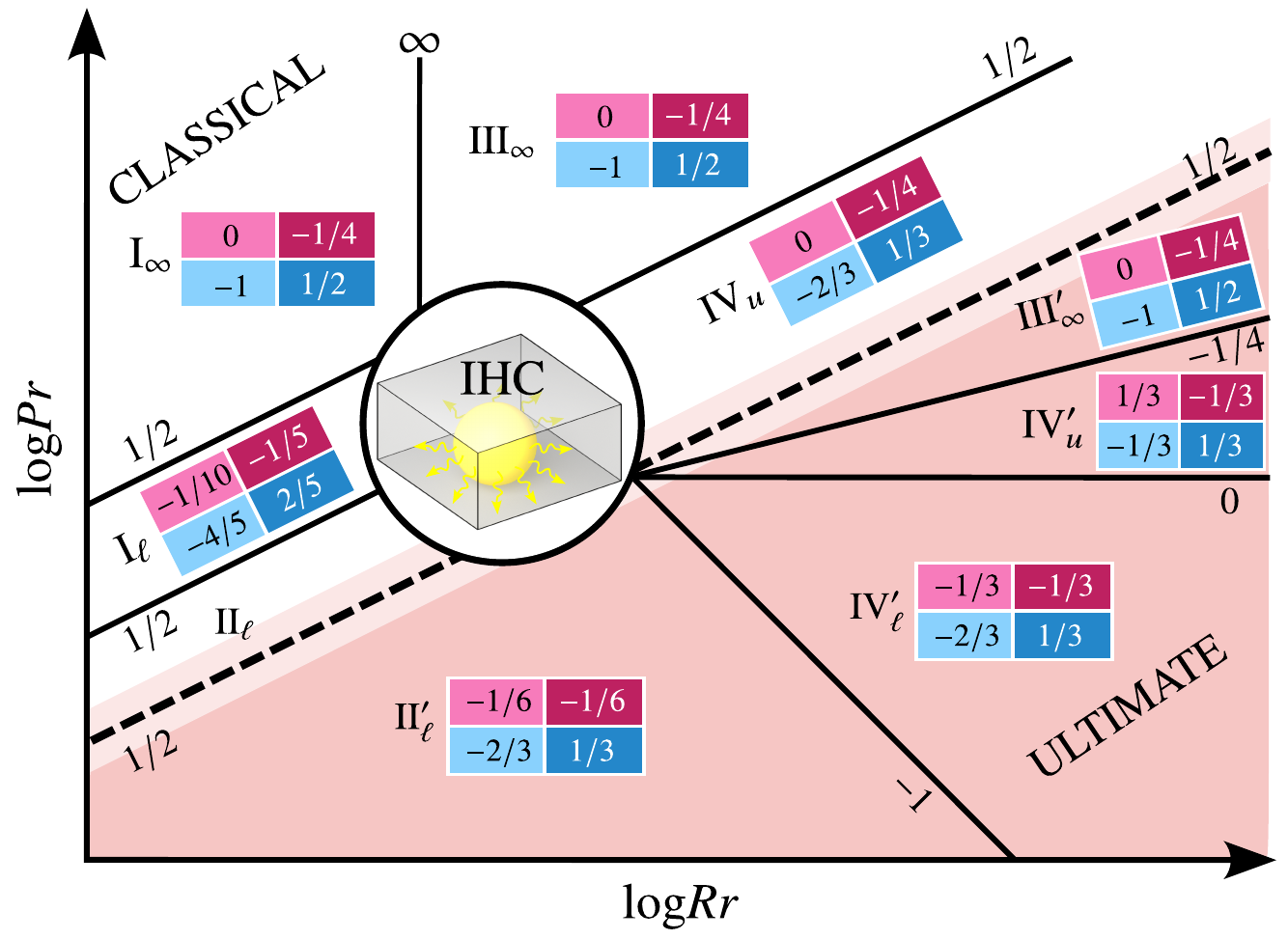}
\put(-240,160){{\fontsize{10pt}{12pt}\selectfont(\textit{c})}}
\end{subfigure}
\vskip-2mm
\caption{\it
Sketches of possible scaling laws in (a) RBC, (b) HC and (c) IHC.
The four exponents in each block refer to the pure scaling laws (a, b) $\Nu \sim \Pran^{\gamma_1}\Ra^{\gamma_2}$ and $\Rey \sim \Pran^{\gamma_3}\Ra^{\gamma_4}$, and
(c) $\Dt \sim \Pran^{\gamma_1}\Rr^{\gamma_2}$ and $\Rey \sim \Pran^{\gamma_3}\Rr^{\gamma_4}$.
Here $\gamma_1$ and $\gamma_2$ are given in the first line of each block in pink and $\gamma_3$ and  $\gamma_4$ in the second line in blue.
The numbers next to the straight lines show the exponent $\eta$ for the slopes of the transitions between the neighboring regimes, (a-b) $\Pran\sim\Ra^{\eta}$ and (c) $\Pran\sim\Rr^{\eta}$.
}
\label{fig1}
\end{figure}

The resulting scaling relations for RBC, HC, and pure IHC in the classical and ultimate regimes are sketched in figure~\ref{fig1}.

\section{Discussion and conclusions}
The turbulent boundary-layer relations for RBC \citep{Shishkina2024} can be transferred to HC and pure IHC. The only change is in the global kinetic-energy balance.
In RBC, one has $\epsu \sim (\nu^3/L^4)\Ra\Pran^{-2}\Nu$, whereas in HC and pure IHC one has, at the scaling level, $\epsu \sim (\nu^3/L^4)\Ra_{*}\Pran^{-2}$, with $\Ra_{*}=\Ra$ or $\Rr$.
The absence of the factor $\Nu$ in HC and pure IHC (as compared to RBC) lowers the ultimate-regime exponent from $1/2$ to $1/3$ for fixed $\Pran$.

For HC, the derived exponent $1/3$ is also the one selected by the rigorous upper bound of \citet{Siggers2004} for the relevant boundary conditions.
For pure IHC, the GL-type model of \citet{Wang2021} already showed that pure IHC is the direct analogue of HC once $\Nu$ is replaced by $\Dt^{-1}$ and $\Ra$ by $\Rr$.
The rigorous results of \citet{Arslan2021} for equal-temperature plates concern the convective flux $\Phi$, not $\Dt^{-1}$.
They therefore constrain the prefactor in the kinetic-energy balance, but do not determine the scaling of $\Dt$.
The present $1/3$ exponent for $\Dt^{-1}$ is thus the pure-IHC counterpart of the HC result, under the condition $\Phi=\order{1}$ \citep{Wang2021}.

We also note that pure IHC should not be mixed with internally heated (and cooled) source--sink convection. In the latter case, including the configurations studied by \citet{Kazemi2022}, part or all of the transport can occur directly between heated and cooled regions in the bulk, so $1/2$-type or intermediate exponents are possible. The physics there is therefore different from that of the pure-IHC setup considered here, in which the net produced heat must cross a boundary layer.

In summary, the ultimate regime in HC and pure IHC consists of four subregimes: two ultimate branches with fixed-$\Pran$ exponent $1/3$, and two branches inherited from the GL-type phase diagram.
In this sense, our recent RBC reformulation \citep{Shishkina2024} is not specific to RBC; rather, it provides a general recipe: once the turbulent wall laws are known, the decisive component is the global kinetic-energy balance of the respective convection problem.
\\ \\
{\bf Acknowledgements}.
The authors thank Zhongzhi Yao for the help with the figures and acknowledge the financial support from the German Research Foundation (DFG), grants Sh405/20 and Sh405/22, and European Research Council (ERC) Advanced Grant no. 101094492 MultiMelt.
\\ \\
{\bf Declaration of interests.}
The authors report no conflict of interest.
\\ \\
{\bf Author ORCID}\\
Detlef Lohse, https://orcid.org/0000-0003-4138-2255\\
Olga Shishkina, https://orcid.org/0000-0002-6773-6464

\bibliographystyle{plainnat}
\bibliography{References}

\end{document}